\documentclass[12pt]{spieman}  
\usepackage{amsmath,amsfonts,amssymb}
\usepackage{graphicx}
\usepackage{setspace}
\usepackage{tocloft}
\usepackage{lineno}
\usepackage{xurl}

\newcommand\comblue[1]{#1}

\def\textmu{${\mu}$m }

\title{3D Printed Alumina as a Millimeter-Wave Optical Element}

\author[1]{Rex Lam}
\author[1]{Scott Cray}
\author[1]{Sam Dietterich}
\author[1] {Calvin Firth}
\author[1] {Shaul Hanany}
\author[2] {Takumi Izawa}
\author[4] {J\"{u}rgen Koch}
\author[3] {Kuniaki Konishi}
\author[2] {Tomotake Matsumura}
\author[3] {Haruyuki Sakurai}
\author[2,6] {Yuki Sakurai}
\author[5] {Ryota Takaku}
\author[1] {Andrew Yan}

\affil[1]{School of Physics and Astronomy, University of Minnesota, Twin Cities, 115 Union St. SE, Minneapolis MN 55455, USA}
\affil[2]{Kavli Institute for the Physics and Mathematics of the Universe (IPMU), The University of Tokyo, 5-1-5 Kashiwa-no-Ha, Kashiwa, Chiba 277-8583, Japan}
\affil[3]{Institute for Photon Science and Technology (IPST), The University of Tokyo, 7-3-1 Hongo, Bunkyo-ku, Tokyo 113-8654, Japan }
\affil[4]{ Laser Zentrum Hannover e.V., Hollerithallee 8, 30419 Hannover, Germany }
\affil[5]{Okayama University, 3-1-1 Tsushimanaka Kita-ku, Okayama, Japan}
\affil[6]{ Suwa University of Science, 5000-1 Toyohira, Chino-shi, Nagano 391-0292, Japan 
}

\cftpagenumbersoff{figure}
\cftpagenumbersoff{table} 
\begin{document} 
\maketitle

\begin{abstract} 
We present millimeter and sub-millimeter \comblue{room temperature} transmission and loss measurements of 3D printed alumina disc and of a disc with one-sided 3D printed sub-wavelength structures anti-reflection coatings (SWS-ARC). \comblue{For four bands spanning 158 - 700~GHz we find an index of refraction consistent with $n= 3.107 \pm 0.007$}.
The loss over the entire frequency band between \comblue{158~GHz and 700~GHz spans $ 1 \cdot 10^{-3} \leq \tan \delta \leq 2.49 \cdot 10^{-3}$} with 10\%-30\% uncertainty at the lower range of frequencies shrinking to $\sim\!2\%$ at the higher frequencies.  As expected, constructive and destructive interference fringes that are apparent with the flat disc data are absent with the disc that has SWS-ARC. The measured data are consistent with finite element analysis predictions that are based on the measured shape of the SWS. At frequencies between \comblue{158}~GHz and 200~GHz, below the onset of diffraction effects, reflectance is reduced from a maximum of 64\% to about 25\%, closely matching predictions. These measurements of the index, loss, and SWS-ARC of 3D printed alumina suggest that the material and fabrication technique could be useful for astrophysical applications. 

\end{abstract}



\section{INTRODUCTION}
\label{sec:intro}  

Alumina is a material widely used in millimeter and sub-millimeter wave astrophysics for lenses and sub-millimeter filters~\cite{mustang2,nadolski2018,bicep2022,sakaguri2022broadband,golec22,Inoue2016}. It has among the lowest loss in the millimeter wave band, relatively high absorption in the IR, and high thermal conductance leading to low emission when heat sunk at cryogenic temperatures~\cite{Inoue:14}. This makes it useful as a low-pass filter for use in \comblue{cryogenic receivers}~\cite{mustang2,bicep2022,golec22}. 

Alumina's high index of refraction $n \simeq 3$~\cite{Lamb1996} requires implementing an anti-reflection coating to avoid large reflections, and several ARC techniques have been proposed.~\cite{nadolski2018,2021PhDTGroh,Priyadarshini2016,Askar2013,Chao2011,canning21}. We have been focused on implementing the technique of sub-wavelength structures (SWS) as an ARC~\cite{rytov1956,Raut2011,mustang2,matsumura2016,wen2021,takakuJAP2020,Schutz2016}. 


Two methods have been proposed to fabricate SWS on alumina: laser ablation~\cite{Oktem2013,Wang2018,mustang2,Schutz2016,matsumura2016} and dicing~\cite{golec2020}. With dicing, Golec et al. successfully fabricated 40 - 50~cm diameter alumina-based filters with pass-bands between 75 and 300~GHz.~\cite{golec2020} A 42~cm diameter filter took 15 days to make. They report that due to the hardness of alumina multiple replacements of dicing blades were required.~\cite{golec2020} Laser ablation is wear-free, and therefore scalable to large quantities and diameters if ablation rates are demonstrated to be sufficiently high.   
To date, authors report ablation rates on alumina between 0.5 and 34~mm$^3$/min~\cite{Furmanski2007, Perrie2005, Beausoleil2020,wen2021}, spanning a variety of ablation conditions including laser type and power, pulse duration, and scan speed. 
The largest alumina optical element that has been laser ablated is 300~mm diameter \comblue{filter} operating with the MUSTANG2 instrument \comblue{at a band between 75 and 105~GHz}.~\cite{mustang2} It took less than 4 days to make, and the reported ablation rate was 18 mm$^{3}$/min.~\cite{mustang2}. 



Another method to fabricate alumina is through additive manufacturing, also known as 3D printing. Additive manufacturing opens pathways for patterning SWS on either flat or curved alumina samples at the time of optical element fabrication obviating the need for subtractive manufacturing. 3D printing might also give higher aspect ratio SWS. Jim\'{e}nez-s\'{a}ez et al.\ measured the room temperature loss of 3D printed alumina near 100~GHz finding a value of $\tan \delta = 3.8 \cdot 10^{-4}$.~\cite{jimenez2019} 

In this paper we report the properties of samples of 3D printed alumina without and with SWS-ARC. We describe the samples in Section \ref{sec:sample}, and the transmission and reflection measurements between \comblue{158 and 700~GHz} of a bare 3D printed alumina and a sample that had SWS 3D printed on one side in Section~\ref{sec:measurements}. The analysis and results are presented in Section~\ref{sec:results}, and we discuss the results and summarize in Section~\ref{sec:discussion}. 

This paper is an update of an earlier conference proceedings publication~\cite{lam2024}. The main updates are the following: (1) we update the analysis of the measured reflectance and transmittance data giving revised index of refraction and loss tangent values; (2) we extract a 3-dimensional average unit cell of the SWS-ARC and use it with an electromagnetic finite element analysis (FEA) software to predict the transmittance and reflectance of the sample; and (3) compare the prediction and measured data. To our knowledge, the conference proceedings publication~\cite{lam2024} and this paper are the first to present 3D printing as a method to implement SWS-ARC with alumina and to give the first measurements of 3D printed alumina over this frequency range. 






\section{Samples}
\label{sec:sample}

\begin{figure}[!htbp]
\begin{center}
\includegraphics[width=5in]{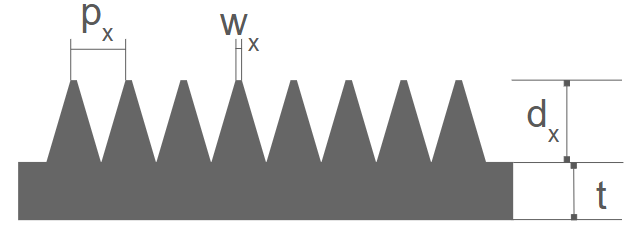}
\end{center}
\caption{Schematic of a 1D cross section in $x$ of the patterned disc. The schematic is not-to-scale. \label{fig:sws_schematic}} 
\end{figure}

\begin{figure}[!htbp]
\begin{center}
\includegraphics[width=1.7in]{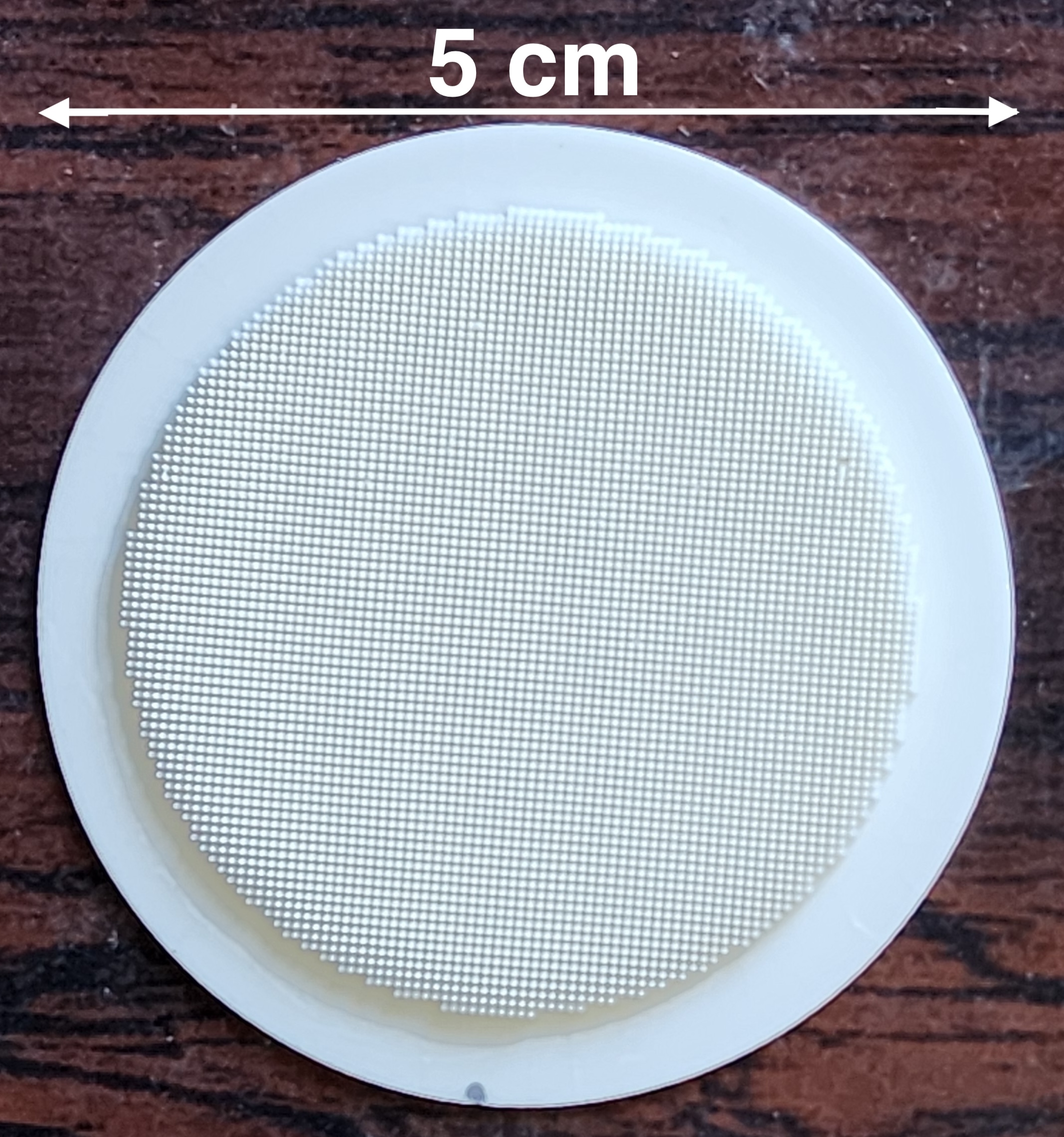}
\includegraphics[width=2.6in]{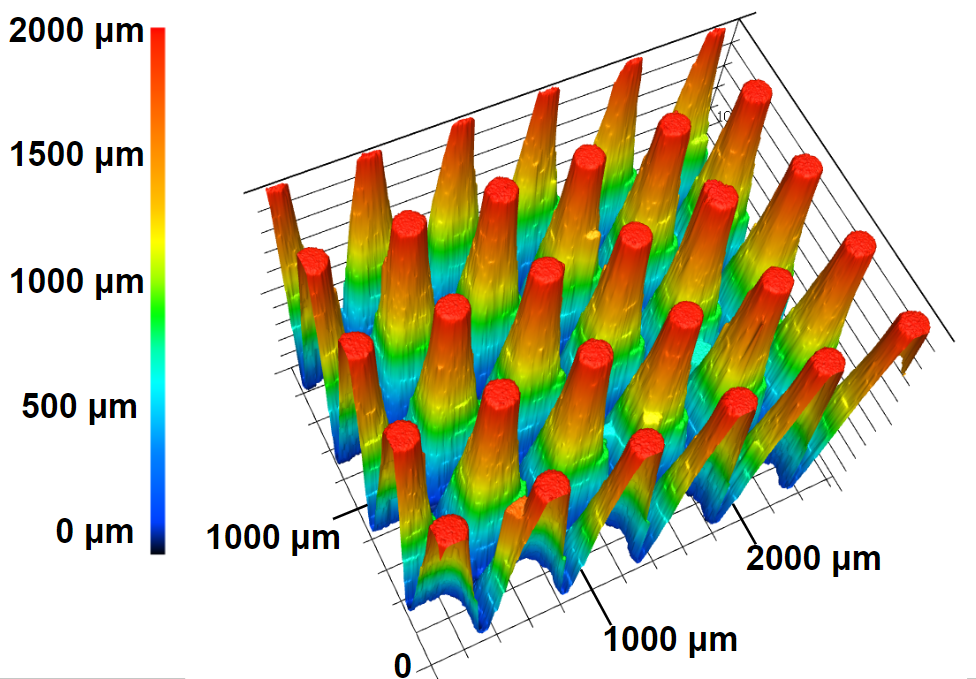}
\includegraphics[width=1.9in]{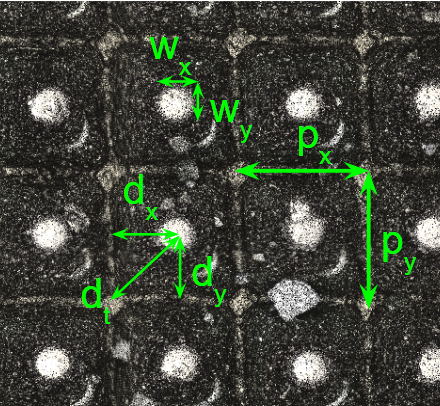}
\end{center}
\caption{\comblue{A photograph of the 3D printed sample (left)}, a confocal microscope images of a section of the sample (middle), and definitions of the shape parameters (right). In the right panel, the ${\rm w}$ and ${\rm p}$ quantities give width and pitch, respectively, and the ${\rm d}$ quantities give depth measured from the tip in the directions shown. 
    \label{fig:scans} }
\end{figure}

Nishimura\footnote{https://nishimuraac.com/}, a company that specializes in sintering alumina powder, partnered with SK FINE\footnote{https://www.sk-fine.co.jp/en/}, a company that specializes in 3D printing, to make two \comblue{99.6\% purity} alumina discs. 
\comblue{The material is mixed, 3D printed, cured, and sintered to produce hard samples similar to standard sintered alumina. The density was measured at 3.9~g/cm$^{3}$.} 
One disc, henceforth referred to as `flat', was 50~mm in diameter and was flat on both sides. \comblue{Its thickness was measured with a micrometer in 10 locations and was found to be $2.940 \pm 0.007$~mm (average $\pm$ standard deviation).} A second disc, also 50~mm in diameter and henceforth referred to as `patterned', had SWS on one side. A schematic cross section of the patterned sample is shown in Figure~\ref{fig:sws_schematic} \comblue{and a photograph is included in Figure~\ref{fig:scans}}.

\comblue{Due to the exploratory nature of this nascent technology, we did not optimize the SWS parameters for a specific instrument or a pre-determined frequency band. Rather, we specified parameters that are typical for instruments operating near a wavelength of 2~mm (150~GHz), which is near the peak in the spectrum of the cosmic microwave background anisotropy. For similar reasons we didn't require that the sample be structured on both sides. As a preliminary assessment of the technology, we deemed a one-sided patterned sample to be sufficient. 
We requested a pitch $p_{x} \simeq 0.5$~mm height $d_{x} \simeq 2$~mm, and tip width that is about twice the resolution of the printing $w_{x} \simeq 0.160$~mm.}

We used a Keyence VKX-3000 confocal microscope to measure the SWS parameters of 12 pyramids. A sample image is given in Figure~\ref{fig:scans} and averages and standard deviations of the measurements are given in Table~\ref{tab:pa_params}. The horizontal $(x,y)$ resolution of the images is 1.3~\textmu and 5~nm in the vertical $z$ direction. The analysis program used to measure the various parameters is outlined in Hanany et al. 2025\cite{lenspaper}. The same program also produces a measured average unit cell of the periodic structure, \comblue{ to which we henceforth refer as the `average pyramid'. }

\begin{table}[!htbp]
    \centering
    \caption{Parameters of 3D printed alumina with patterned SWS. The parameters are defined in Figure \ref{fig:sws_schematic} and the right panel of Figure \ref{fig:scans}. The data give averages and standard deviations over 12 pyramids.
    \label{tab:pa_params} }
    \begin{tabular}{c c } 
        Parameter &  Measurement (mm)  \\ \hline
        Thickness of Substrate ($\rm{t}$) & $1.04 \pm 0.01$ \\
        Pitch x ($\rm{p_x}$) & $0.490 \pm 0.004$  \\
        Pitch y ($\rm{p_y}$) & $0.489 \pm 0.003$ \\
        Tip Width x ($\rm{w_x}$) & $0.176 \pm 0.004 $  \\
        Tip Width y ($\rm{w_y}$) & $0.164 \pm 0.010 $  \\
        Saddle depth x ($\rm{d_x}$) & $1.92 \pm 0.02$ \\
        Saddle depth y ($\rm{d_y}$) & $1.92 \pm 0.03$ \\
        Total Depth ($\rm{d_t}$) & $1.955 \pm 0.006$   \\
        \hline
        \end{tabular}
\end{table}


\section{Measurements} 
\label{sec:measurements}

\begin{figure}[!htbp]
\begin{center}
\includegraphics[width=2.5in]{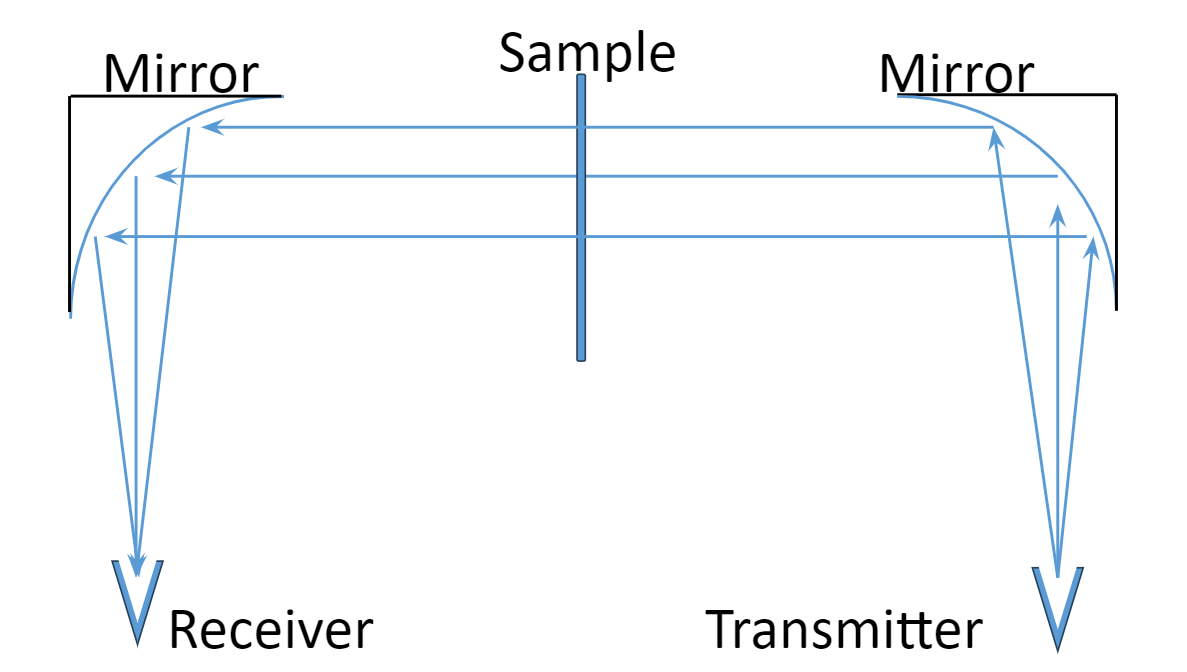}
\hspace{0.1in}
\includegraphics[width=2.5in]{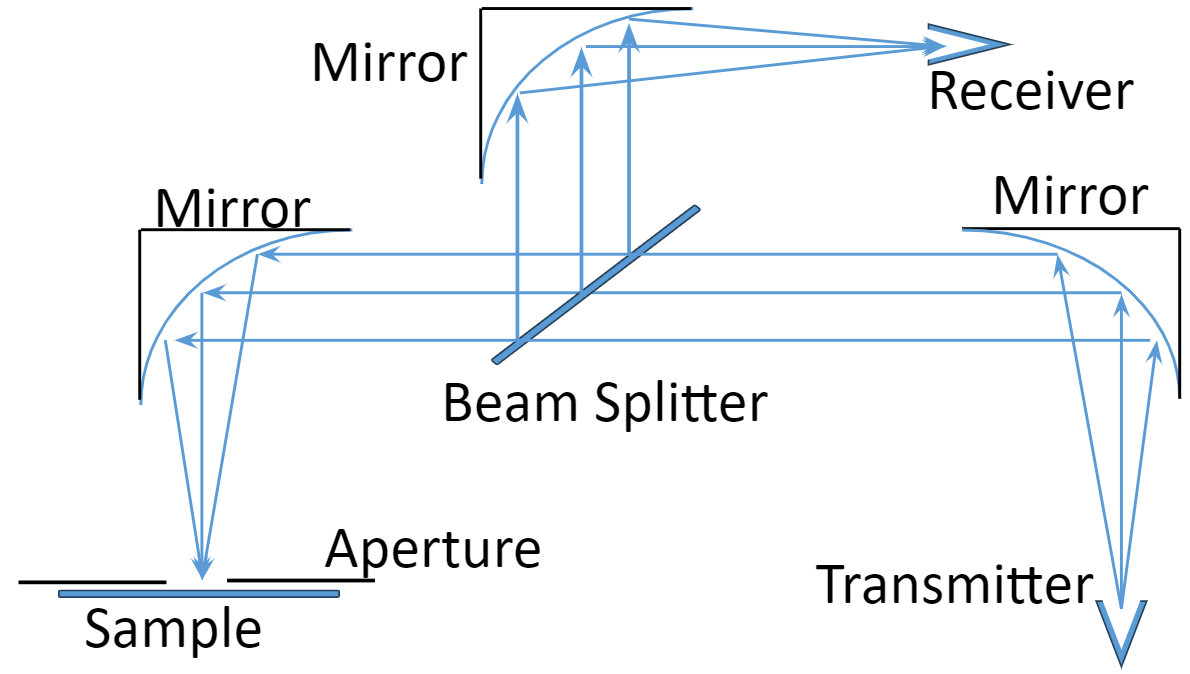}
\end{center}
\caption{Experimental configurations for transmission (left panel) and reflection measurements (right panel). The `Transmitter' and `Receiver' functions are provided by a vector network analyzer. The sketches are not to-scale.  
    \label{fig:setups} }
\end{figure}

Transmission and reflection measurements were performed for both samples using a vector network analyzer (VNA) and the experimental setups depicted in Figure~\ref{fig:setups}. Each measurement was repeated twice to assess reproducibility and the results were found to be identical to within a percent. 
Therefore, only a single set of measurements was used for subsequent analysis. The transmission and reflection of the flat disc was measured between \comblue{158 and 700~GHz} with four sub-bands as listed in Table~\ref{tab:flat_params}.
The patterned disc was measured only \comblue{between 158 and 230~GHz} because diffraction effects were expected to dominate measurements at higher bands; see Section~\ref{sec:discussion}.



The transmission measurement is normalized by a measurement without the sample and the reflection measurement is normalized by comparing the measurement to a measurement with a gold mirror in place of where the sample would be. Henceforth, unless otherwise stated explicitly, the `transmission' or `reflection measurements' refer to the normalized version of the data, which are ratios of electric field amplitudes. We use the terminology $T\!\!=$`transmittance' and $R\!=$`reflectance' to refer to ratios of electric field squared values. If $A\!\!=$Absorptance then $T+R+A=1.$

\section{Analysis and Results} 
\label{sec:results}

\subsection{Outlier removal}
\label{sec:outlier}

The transmission and reflection data are sensitive to spuriously low counts in the normalization data and to other systematic effects which produce data outliers. We identify and remove outliers with a three-step process \comblue{that is applied to the flat sample}. First, \comblue{we remove points} that have a transmission value above~1.3. Transmission values above 1 are unphysical. We best fit the remaining transmission data to a transmission model constructed using a transfer matrix method (TMM)~\cite{hecht} parametrized by the known thickness of the sample and unknown index of refraction and loss tangent values. We subtract \comblue{the best fit from the data} to form a residual, bin the residual, fit it with a Gaussian and find the mean and standard deviation $\sigma$. We reject all data points that are $3\sigma$ from the mean. This process is repeated a second time. In total 
\comblue{1.6\%, 1.2\%, 1.1\%, and 1.8\%}
of the points are removed per frequency band, enumerated from lower to higher frequencies. 


\subsection{Flat Disc}
\label{sec:flat}

\begin{figure}[!htbp]
\begin{center}
\includegraphics[width=6.5in]{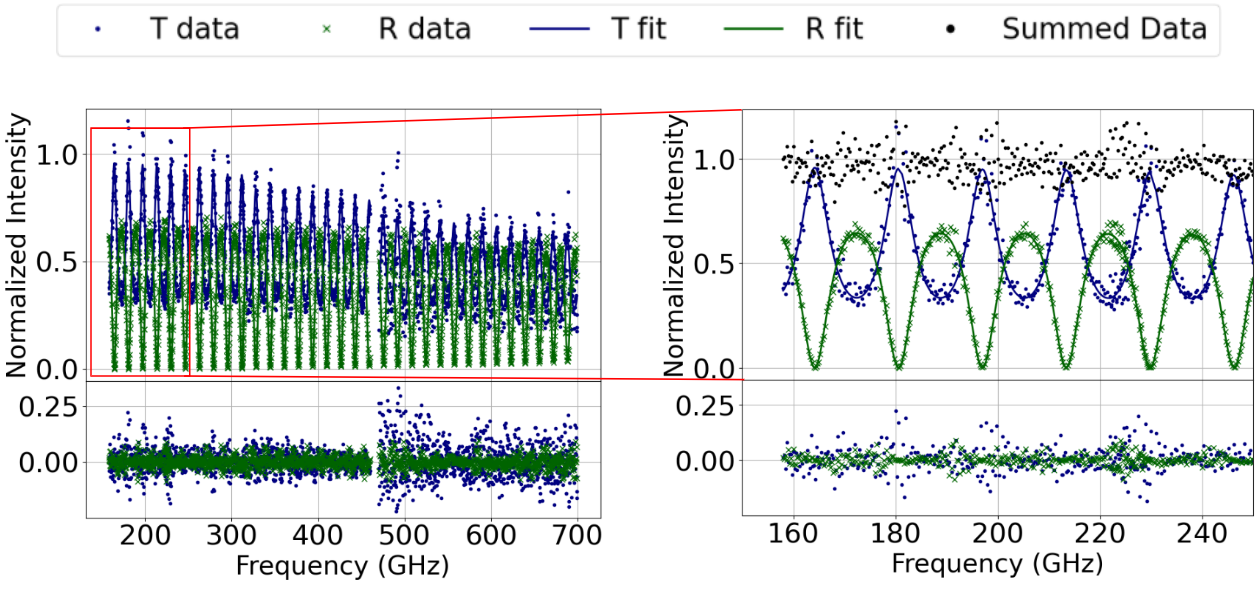}
\end{center}
\caption{Transmittance $T$ and reflectance $R$ data (upper panels, blue and green points, respectively) as a function of frequency for the flat disc, the best fit models (solid blue and green, respectively), and the residuals from the best fits (lower panels). The expanded view of a low frequency band (right panel) also gives the sum of the measured transmittance and reflectance (black).
    \label{fig:flatdata_and_fits} }
\end{figure}

The transmission and reflection data measured with the flat disc are used to extract the index of refraction $n$ and loss tangent $\tan \delta$ of the 3D printed material. 
We least-squares fit the transmission and reflection data to a TMM model that includes the measured thickness of the sample. This is the same model used in Section~\ref{sec:outlier} and the extracted parameters $n$ and $\tan \delta$ are those obtained from the best fit after outlier rejection. 
Each data point is assigned an uncertainty equal to the standard deviation $\sigma_{x}$ of the data residual where data residual is equal to the data minus the best fit, and $x=t\, \mbox{or}\, r$ refers to using either transmission or reflection data. 
We find a pair of $n$ and $\tan \delta$ for each frequency band.  
The reflectance and transmittance data and the best fits are displayed in Figure~\ref{fig:flatdata_and_fits}.
The uncertainties in the parameters are determined using Monte Carlo simulations~\cite{numericalrecipes} with 500 mock data sets. Mock transmission and reflection data points are assumed to be Gaussian distributed around the original data with $\sigma_{x}, \, x=t \, \mbox{and}\, r$. From these mocks we find the distribution of parameter values. For the index $n$ we find that the $1\sigma$ statistical uncertainty, which is less than 0.02\%, is dwarfed by a 0.24\% systematic uncertainty arising from uncertainty in the thickness of the sample. The thickness uncertainty leads to an uncertainly $\Delta n = 0.007$ at all frequency bands; see Table~\ref{tab:flat_params}. The thickness uncertainty plays a smaller role for $\tan \delta$ giving values $\Delta \tan \delta$ that are comparable to the statistical uncertainties. The final $\Delta \tan \delta$ we report in Table~\ref{tab:flat_params} is the quadrature sum of the statistical and systematic uncertainties. 


\begin{table}[!htbp]
    \centering
    \caption{ The index of refraction and loss tangent values and 68\% confidence intervals for the flat sample for each frequency band. 
    \label{tab:flat_params} }
    \begin{tabular}{c| c | c  } 
        Frequency band (GHz) &  Index $n$ & Loss $\tan \delta \, (\times 10^{3})$ 
        \\ \hline
       158 - 230 & $3.106 \pm 0.007 $ & $1.0 \pm 0.1$ \\
%
        222 - 315 & $3.106 \pm 0.007 $ & $1.1 \pm 0.3$ \\ 
%
        312 - 460 & $3.108 \pm 0.007 $ & $1.80 \pm 0.03$ \\
%
        470 - 700 & $3.109 \pm 0.007 $ & $2.49 \pm 0.03$ \\       
    \end{tabular}
\end{table}


\subsection{Patterned Disc}

\begin{figure}[!htbp]
\begin{center}
\includegraphics[width=2.1in]{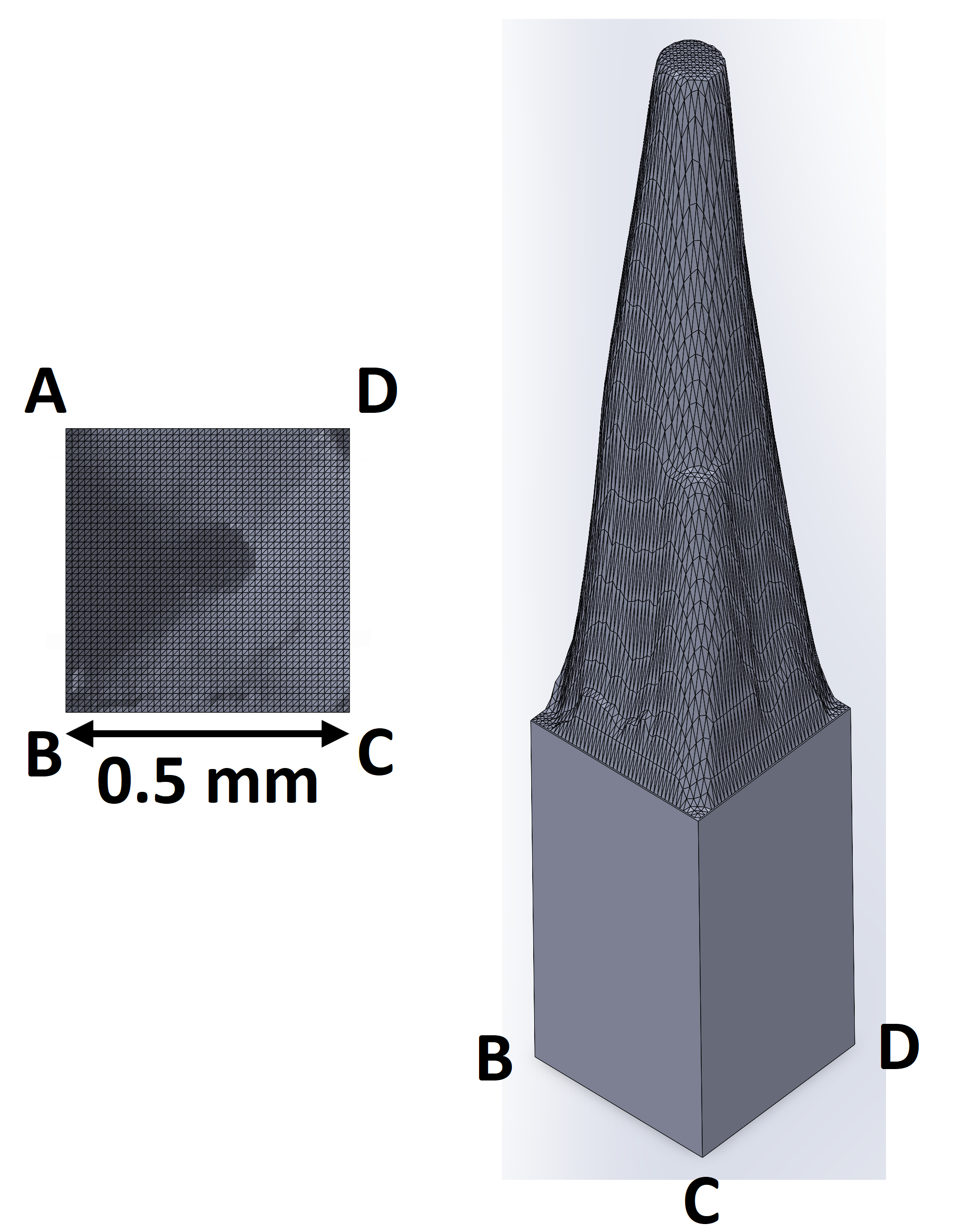}
\includegraphics[width=3.9in]{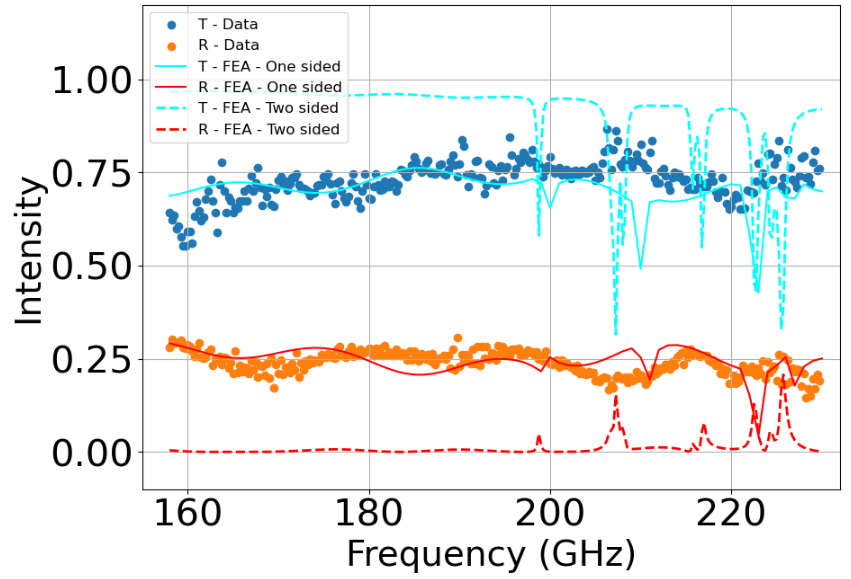}
\end{center}
\caption{\comblue{Top and a perspective view of the unit cell used for FEA calculations (left),} and measured transmittance and reflectance of the patterned disc (right, blue and orange data, respectively) together with FEA generated predictions (right, cyan and red solid). For the FEA predictions we use the unit cell (left), which is constructed from the average pyramid. We also include FEA predictions of performance if both sides of the sample would have been patterned (cyan and red dash). 
\label{fig:patterndata_and_fits} 
}
\end{figure}

We measured transmittance and reflectance with radiation polarized along each of the orthogonal groove directions. The two orthogonally polarized data sets, for both transmittance and reflectance, are consistent within noise. Over all frequencies the average of the differences are 0.022 and -0.010 for transmittance and reflectance, respectively, and the standard deviations are 0.045 and 0.035, respectively. We therefore only show data from one of the polarizations, see Figure~\ref{fig:patterndata_and_fits}. 



We compare the measurements to predictions using an electromagnetic propagation FEA program~\cite{hfss}. We \comblue{generate predictions by constructing a solid model of a unit cell of the patterned disc using the average pyramid, see Figure~\ref{fig:patterndata_and_fits}.} The model is input into the FEA and we assume periodic boundary conditions and the index and $\tan \delta$ that were derived from the flat sample. The FEA-calculated transmittance and reflectance, which were calculated with a 0.25~GHz frequency resolution, are the solid lines in Figure~\ref{fig:patterndata_and_fits}. The sharp features in the predictions are discussed in Section~\ref{sec:discussion}.

\section{Discussion and Summary}
\label{sec:discussion}

\subsection{Index and Loss}

\comblue{Between the lower and higher frequencies we find index values for the 3D printed material between 3.106 and 3.109}. The index uncertainty of 0.007, which is dominated by thickness uncertainty, is larger than the difference between the central values. \comblue{Within uncertainty we quote for the entire band the average index $n=3.107 \pm 0.007$.}   The index values are in agreement with other room temperature measurements of alumina which span a range of $3.05 \leq n \leq 3.145$ between 100 and 400~GHz.~\cite{Lamb1996} For absorptive loss we are finding values between $1.0\cdot 10^{-3}$ and $2.49\cdot10^{-3}$ between 158 and 700~GHz (Table~\ref{tab:flat_params}), and we observe a trend of higher loss at higher frequencies. Both the loss values and the trend are consistent with other reported results giving $ 6 \cdot 10^{-4} \leq \tan \delta \leq 26 \cdot 10^{-4} $ between 100 and 400~GHz.~\cite{Lamb1996} At lower frequencies, both Takaku et~al.~\cite{mustang2} and Jim\'{e}nez-s\'{a}ez~\cite{jimenez2019} report lower loss values -- lower than $5\cdot 10^{-4}$ and $3.8\cdot 10^{-4}$ -- for standard sintered alumina and 3D printed alumina, respectively.

\subsection{ARC Efficacy of 3D Printed SWS-ARC}

The one-sided 3D printed SWS-ARC eliminates the oscillations in transmittance and reflectance that have a magnitude of $\sim\!0.6$ to produce a nearly flat response between 160 and 230~GHz, demonstrating the efficacy of the approach.  The electromagnetic FEA calculation shows the signatures of diffraction in the form of sharp dips in the transmittance and reflectance. 
Given the measured pitch and index of refraction, these are calculated to set in at frequencies near and above 200~GHz,~\cite{takakuJAP2020} and this is the reason we didn't conduct measurements at higher frequency bands. The effects are more pronounced with the simulations, which assume an infinite sample, a uniform zero incidence angle, and neglect the details of a detector horn area and solid angle response. 

To demonstrate the potential impact of SWS-ARC with additive manufacturing we show in Figure~\ref{fig:patterndata_and_fits} the FEA-predicted performance of the sample if both sides were patterned with the average pyramid, see the dashed lines. At frequencies between 150 and 195~GHz, below the first diffraction peak, the average transmittance and reflectance are 96\% and 0.3\%, respectively. The $\sim1.9$~mm height of the structures gives a predicted low reflectance response at even lower frequencies. For example, the calculated average transmittance and reflectance \comblue{between 70 and 195~GHz are 97\% and 0.9\%, respectively. At a frequency of 60~GHz the reflectance rises to 10\%.} Additional measurements in this lower frequency band are left for a future work. 

\comblue{To a first approximation, the start of the pass-band of SWS-ARC is determined by the height of the structures $\rm d$. The larger $\rm d$ the lower the turn-on frequency of the pass-band. The turn-off of the pass-band at the high frequency is determined by the periodicity ${\rm p}$. At normal incidence, diffraction effects, such as those seen in Figure~\ref{fig:patterndata_and_fits}, are expected at frequencies above $\nu \sim c/n{\rm p}$, where $c$ is the speed of light and $n$ is the index of refraction~\cite{RaguinDanielH.1993Aoas}.  The vendor reports a horizontal resolution limit of 80~$\mu$m for the additive manufacturing process. Assuming pyramids with 160~$\mu$m bases, to allow for some structure shaping that converges to 80~\textmu tips, we find an approximate frequency upper limit of $\simeq$600~GHz. There is no rigid constraint on the pass-band turn on at low frequencies because very tall structures can be made with sufficiently broad bases. }

\subsection{Summary}

To our knowledge, the conference proceedings publication~\cite{lam2024} and this paper, which gives a revised and improved analysis, are the first measurements of the optical properties of 3D printed alumina between 160 and 700~GHz, and the first demonstration of 3D printed structures as SWS-ARC. The measurements indicate that the approach is promising, and over time the technology could mature to samples larger than few cm in diameter. 
For future samples, it would be important to \comblue{demonstrate the capability to 3D print on two sides of a sample, to print to pre-specified shape requirements, obtain (or meaure) additional properties such as thermal conductance at various temperatures, and perhaps provide a broader range of material properties, for example, loss tangents, to match a variety of applications.} When mature, 3D printed samples could be used as filters or lenses with astronomical telescopes. 


\section*{ACKNOWLEDGMENTS}       
 
We thank Patrick Camilleri for use of the VKX-3000 confocal microscope with which we imaged the patterned alumina. We thank Nick Agladze at ITST/UC Santa Barbara for the transmission and reflection measurements and for the left panel photograph in Figure~\ref{fig:scans}. Parts of this work were conducted in the Minnesota Nano Center, which is supported by the National Science Foundation through the National Nanotechnology Coordinated Infrastructure (NNCI) under Award Number ECCS-2025124, and
in the Characterization Facility, University of Minnesota, which receives partial support from the NSF through the MRSEC (Award Number DMR-2011401) and the NNCI (Award Number ECCS-2025124) programs. This work was supported by NSF grant numbers NSF-2206087 and NSF-2348668. This work was supported by JSPS KAKENHI Grant Number JP23H00107, and JSPS Core-to-Core Program, A. Advanced Research Networks. This work was performed in part at the Center for Data-Driven Discovery (CD3), Kavli IPMU (WPI). The Kavli IPMU is supported by World Premier International Research Center Initiative (WPI Initiative), MEXT, Japan. This study was funded in parts by MEXT Quantum Leap Flagship Program (MEXT Q-LEAP, Grant Number JPMXS0118067246).

\section*{DISCLOSURES}

The authors declare that there are no financial interests, commercial affiliations, or other potential conflicts of interest that could have influenced the objectivity of this research or the writing of this paper.

\section*{CODE, DATA, AND MATERIALS AVAILABILITY}

The experimental and simulation data used to generate Figure \ref{fig:patterndata_and_fits} are posted on this site: \\ https://conservancy.umn.edu/workflowitems/81826/view .

\newpage

\bibliography{report} 
\bibliographystyle{spiejour} 

\listoffigures
\listoftables

\end{document}